# Relativistic artificial molecules with tunable coupling and orbitals

Xiao-Feng Zhou[1,§], Yu-Chen Zhuang[2,§], Mo-Han Zhang[1], Hao Sheng[1], Qing-Feng Sun[2,3,†] & Lin He[1,†]

**Affiliations:**

[1]Center for Advanced Quantum Studies, Department of Physics, Beijing Normal University, Beijing, 100875, China, and Key Laboratory of Multiscale Spin Physics, Ministry of Education, Beijing, 100875, China

[2]International Center for Quantum Materials, School of Physics, Peking University, Beijing, 100871, China

[3]Collaborative Innovation Center of Quantum Matter, Beijing 100871, China

[§]These authors contributed equally to this work.

[†]Correspondence and requests for materials should be addressed to Qing-Feng Sun (e-mail: sunqf@pku.edu.cn) and Lin He (e-mail: helin@bnu.edu.cn).

**In a molecule formed by two atoms, energy difference between bonding and antibonding orbitals should depend on distance of the two atoms. However, exploring molecular orbitals of two natural atoms with tunable distance has remained an outstanding experimental challenge. Graphene quantum dots (GQDs) can be viewed as relativistic artificial atoms, therefore, offering a unique platform to study molecular physics. Here, through scanning tunneling microscope (STM), we create and directly visualize the formation process of relativistic artificial molecules based on two coupled GQDs with tunable distance. Our study indicates**

**that energy difference between the bonding and antibonding orbitals of the lowest quasibound state increases linearly with inverse distance of the two GQDs due to the relativistic nature of the artificial molecule. For quasibound states with higher orbital momenta, the coupling between these states leads to half-energy spacing of the confined states because the length of the molecular-like orbit is about twice that of the atomic-like orbit. Evolution from ring-like whispering-gallery modes in the artificial atoms to figure-eight orbitals in the artificial molecules is directly imaged. The ability to resolve the coupling and orbitals of the relativistic artificial molecule at the nanoscale level yields insights into the behavior of quantum-relativistic matter.**

Quantum dots (QDs) can be viewed as artificial atoms due to their abilities to confine electrons into atomic-like discrete energy levels[1–5]. With reducing the distance, orbitals of two coupled QDs can linearly combine to form two new molecular-like states, *i.e.*, the bonding and antibonding molecular orbitals, thus creating artificial molecules[6-30]. Artificial molecules not only provide significant advantages in the study of atomic and molecular physics, but also play a vital role in the fabrication of quantum devices. Therefore, many efforts have been devoted to fabricate and study the artificial molecules[7–24]. However, systematic study of the evolution from the artificial atoms to the artificial molecules is still lacking owning to the difficulty to achieve continuous and precise control of the coupling strength.

Here, by applying scanning tunneling microscope (STM) tip pulses to a graphene/WSe$_2$ heterostructure, we create two almost identical graphene/WSe$_2$ heterostructure quantum dots (GQDs), and continuously tune the distance between the two GQDs with nanoscale precision. In this process, the two GQDs evolve from two artificial atoms to an artificial molecule with clear bonding and antibonding states. Our experiment supported by theoretical calculations indicates that the energy spacing of the bonding and antibonding states formed by the lowest quasibound state varies linearly with inverse of the distance. This phenomenon is unique to relativistic artificial molecules and is different from that predicted in semiconductor artificial molecules[7,9,11,25–30]. For quasibound states with higher orbital momenta, we observe figure-eight molecular orbits and repulsive circular orbits from two coupled ring-like whispering-gallery modes (WGMs)[24]. The energy spacing between them is half of that in a single GQD, which arises from the fact that the length of the molecular-like orbit is about twice that of the atomic-like orbit. Spatial distributions of the molecular orbitals in the artificial molecule with custom-designed coupling are directly imaged by using STM and they are also reproduced well by our theoretical calculations.

In our experiment, high-quality graphene/WSe$_2$ heterostructure was obtained by wet transfer technology of a graphene monolayer on mechanical-exfoliated thick 2H-phase WSe$_2$ sheets[23,31–33] (see Methods section for details of fabrication). Figure 1a shows schematic of the strategy for fabricating and manipulating the GQDs. As reported previously, a nanoscale pit and a 1T'-phase monolayer WSe$_2$ island can be created at the interface of the graphene/WSe$_2$ heterostructure by using a STM tip pulse[33].

The different doping levels of the 2H-phase and 1T'-phase WSe$_2$ to the supported graphene introduce a sharp electronic junction, which generates a nanoscale GQD in the continuous graphene monolayer. By applying pulses of the same parameters to the STM tip, we created two almost identical GQDs (see Fig. S1 for STM characterizations of the two GQDs). In order to reduce the impact of the pits on the electrical properties of the GQDs, the two GQDs are moved to a flat area away from the pits. Then, the distance $d$ between the two GQDs is tuned continuously with nanoscale precision to change the coupling strength between them (see Supplementary Section 2 for details of manipulation). Figure 1b shows a representative STM image of two GQDs with a distance $d \approx 27$ nm and the profile line across the GQDs indicates that they are generated by two WSe$_2$ monolayer islands. Figure 1c shows the scanning tunneling spectroscopy (STS), *i.e.*, d$I$/d$V$, map measured across the centers of the two GQDs indicated by the red line in Fig. 1b. Obviously, the confined potential introduced by the 1T'-phase WSe$_2$ islands generates a series of quasibound states in the GQDs. Because of the similar size of the two GQDs, the energies of the quasibound states in each GQD are almost the same, which is necessary for them to couple to form the bonding and antibonding orbitals according to molecular orbital theory[6]. However, at this distance ~ 27 nm, the coupling between the two GQDs is very weak and the system can be regarded as two isolated GQDs approximately. To explicitly show this, we carried out the energy-fixed d$I$/d$V$ mappings, which reflect the spatial distributions of the LDOS, at the N1 of the two GQDs, as shown in Figs. 1d and 1e. Due to the two GQDs being almost identical, their levels are approximately degenerate and the spatial distributions of the N1 of the dot 1 is almost the same as that of the dot 2. Obviously, the electronic states of the N1 are mainly localized in the centers of each GQD, which is similar as that of an isolated GQD, further confirming that there is no significant coupling between the two GQDs. To fully understand the observed quasibound states, the confined potential along one of the GQDs is measured according to the spatial dependence of the Dirac point $E_D$ (see Fig. S1) and it can be approximately described by an electrostatic potential as follows:

$$V_s(r) = \begin{cases} V_{in}, & r < r_{in} \\ V_{in} - k(r - r_{in}) & r_{in} < r < r_{out} \\ V_{out}, & r > r_{out} \end{cases}, \qquad (1)$$

where $r$ is the distance from the center of GQD, $r_{in}$ is the radius of the GQD, and $V_{in}$ and $V_{out}$ are the constant electrostatic potentials inside and outside the GQD due to the screening effect (see Supplementary Section 3 for more details of $V_s(r)$). Our theoretical calculation based on the confined potential well reproduces the main features of the quasibound states observed in our experiment and confirms that the observed quasibound states are formed via the WGMs in an isolated GQD[19,23,31–41] (see Supplementary Fig. S1). Therefore, the lowest quasibound state exhibits a maximum in the center of the GQD and higher-energy quasibound states with higher angular momenta display ring-like structures along the edge of the GQD.

To explore molecular states formed by the two GQDs, we systematically decreased the distance between the two GQDs and studied their quasibound states. By moving one GQD to the other through STM tip manipulation, twelve different distances ranging from 30 nm to 12 nm between the two GQDs are set in our experiment (see Fig. S2 for STM images of the two GQDs with different distances). Figure 2 shows representative results of the two GQDs with three different distances. For the distance $d > 21$ nm, as shown in Fig. 1, the coupling between the two GQDs is very weak and the quasibound states of the two GQDs are almost the same as that of the isolated GQD. By decreasing the distance to $21 \text{ nm} > d > 14 \text{ nm}$, the lowest quasibound states of the two GQDs are slightly split, as shown in Figs. 2b and 2e. Such a result reminds us the formation of the molecular states[19,24]. Then, the two states split from the lowest quasibound states N1 of the two GQDs can be viewed as the bonding (N1+) and antibonding (N1-) states of the newly formed artificial molecule[6-30]. The same conclusion can also be obtained from the energy-fixed $dI/dV$ mappings. Figures 2c and 2f show the LDOS distributions of the two GQDs that are attractive at the bonding state and repulsive at the antibonding state, which are consistent with the formation of the molecular states. By decreasing the distance from 18 nm to 15 nm, the potential fields of the two GQDs are strongly overlapped, as indicated by the black dashed line in Fig. 2e. Then, the bonding and antibonding states become more pronounced and the energy splitting between them becomes larger, as shown in Figs. 2e and 2f. By further decreasing the distance to $d < 14$ nm, two GQDs are almost in contact. In this case, the potential fields of them are translated from two circular GQDs into a single elliptical-like GQD, as shown in Figs. 2h and 2i (see Figs. S3 and S4 for more $-d^3I/dV^3$ STS maps and energy-fixed $dI/dV$ mappings with other distances, respectively). So far, we have realized the evolution from two atomic-like states into molecular-like states in two GQDs through tip manipulation.

To fully understand our experimental results, we calculated the evolution of the electronic properties of the two GQDs (see Methods section for the model of theoretical simulations). To describe the coupling between two GQDs, we set a molecular potential field $V_m(r_1, r_2)$, which represents the superposition of the potential fields of two GQDs. Due to the fixed doping level of the 1T' WSe$_2$ on graphene, the maximum value of $V_m(r_1, r_2)$ should not exceed the potential in the single GQD ($V_{in}$), therefore, we cut the value of $V_m(r_1, r_2)$ as $V_{in}$ once it is larger than $V_{in}$. That is, the molecular potential field $V_m(r_1, r_2)$ is set as:

$$V_m(r_1, r_2) = \begin{cases} V_s(r_1) + V_s(r_2) & V_s(r_1) + V_s(r_2) \leq V_{in} \\ V_{in} & V_s(r_1) + V_s(r_2) > V_{in} \end{cases}, \quad (2)$$

where $r_1$ and $r_2$ are the distances from the centers of the dot 1 and dot 2, respectively. In the numerical simulation, the two GQDs are set at x axis with the inter-dot distance $d$. Figures 3a-3f show the calculated LDOS space-energy maps and the corresponding energy-fixed LDOS mappings of the bonding and antibonding states at the same distances as in Fig. 2 (see Figs. S5 and S6 for more LDOS space-energy maps and energy-fixed LDOS mappings with other distances, respectively). Obviously, they are in good agreement with the experimental results, thus further supporting our qualitative understanding. Both our experiments and theoretical simulations indicate that the coupling between the two GQDs becomes stronger with decreasing the distance. To quantitatively show this, Figure 3g summarizes the distance-dependent splitting ($\Delta E$) of the lowest quasibound states ($N_{1\pm}$) of the two GQDs obtained by both experiment and theoretical simulation. The obtained $\Delta E$ in the experiment and theory are quite similar and both results show that the relation between $\Delta E$ and $1/d$ is approximately linear (see Supplementary Section 9 for the parameters of the linear fits in Fig. 3g).

Theoretically, the bonding and antibonding states of the artificial molecules are formed by the overlapping of the electron wave functions of the artificial atoms, and they can be also viewed as the quasibound states of the new molecular potential field with the effective confined size that is approximately proportion to the inter-dot distance $d$, as shown in the inset of Fig. 3g. Combining with the linear dispersion relation of massless Dirac fermions in graphene with Fermi velocity $v_F$, the energy spacing of the bonding and antibonding states $\Delta E$ is approximatively proportional to $\hbar v_F/d$. Therefore, this phenomenon underscores the uniqueness of the relativistic artificial molecules compared with conventional molecules realized by coupled natural atoms or semiconductor QDs. The slight deviations of the slope visible between experiment and simulation is probably due to the deviation in the experimental Fermi velocity from $v_F = 1.03 \times 10^6 \ m/s$ used in our simulation (see methods). The slight differences in the theoretical and experimental potential fields of the GQDs may also weakly affect the $\Delta E$.

Besides the lowest quasibound states, the other quasibound states with higher angular momenta in the two GQDs can also couple to form the molecular orbits. Then, the two ring-like structures of the two GQDs are attractive in the bonding state and repulsive in the antibonding state, as schematically shown in Fig. 4a. The formation of the molecular orbits by the quasibound states with higher angular momenta is clearly shown in the STS maps measured across the centers of the two GQDs (Figs. 2b, 2e, and 2h), which is more evident in the zoomed-in view, as shown in Fig. 4b as an example for result of the two GQDs with $d \approx 12$ nm. Obviously, the coupling of the two GQDs strongly modifies the quasibound states with higher angular momenta, which exhibit a distinct behavior compared with that in an isolated GQD (see Fig. S7 for the energy-fixed $dI/dV$ mappings of the uncoupled quasibound states). In Fig. 4b, we can distinguish the bonding state $N_{2+}$ and antibonding state $N_{2-}$ arising from two ring-like atomic orbits $N_2$ in Fig. 1c. In view of the strong coupling, they are also partially influenced by overlapping with quasibound states of other energies. The spatial distributions of the molecular orbits formed by the quasibound states with higher angular momenta can be explicitly visualized in the STS maps, as shown in Figs. 4c and 4d. The bonding state exhibits a characteristic figure-eight orbital and the antibounding state exhibits two repulsive circular orbitals. Such a result is also reproduced well in our theoretical simulations with considering two coupled GQDs, as shown in Figs.4e-4g (see Fig. S8 for more calculated energy-fixed LDOS mappings with other distances). A direct and pronounced result of the formation of the molecular orbits by the quasibound states with higher angular momenta is that the energy spacing of the quasibound states is changed to half of that in an isolated GQD. The half-energy spacing observed in the molecular-like states compared with the atomic-like states is due to the length of the molecular-like orbit being twice that of the single atomic-like orbit. According to the semiclassical quantization rule, the energy spacing between the quasibound states is proportional to $\hbar v_F/L$, where $L$ is the semiclassical orbit length.

In summary, we create and systematically study the relativistic artificial molecules based on two coupled GQDs with tunable distance. Our experiment reveals that the energy spacing of the bonding and antibonding states formed by the lowest quasibound state increases linearly with inverse of the distance because of the relativistic nature of the artificial molecule. The formation of molecular states by quasibound states with higher orbital momenta leads to half-energy spacing of the confined states due to the twice length of the molecular-like orbit comparing to that of the atomic-like orbit. Our study, especially the direct visualization of the molecular orbits, lays a solid foundation for understanding the relativistic molecular physics and provides an approach that is generally applicable for realizing patterns of coupled GQDs with tunable coupling strength.

**Data availability**

All data supporting the findings of this study are available from the corresponding author upon reasonable request.

**Acknowledgments**

This work was supported by the National Key R and D Program of China (Grant Nos. 2021YFA1400100 and 2021YFA1401900), National Natural Science Foundation of China (Grant Nos. 12141401, 11974050, 11921005, 12374034), "the Fundamental Research Funds for the Central Universities" (Grant No. 310400209521), the Innovation Program for Quantum Science and Technology (2021ZD0302403), and the Strategic priority Research Program of Chinese Academy of Sciences (Grant No. XDB28000000). The devices are fabricated from Shanghai Onway Technology Co., Ltd..

**Author contributions**

X.F.Z. performed the sample synthesis, characterization and STM/STS measurements.

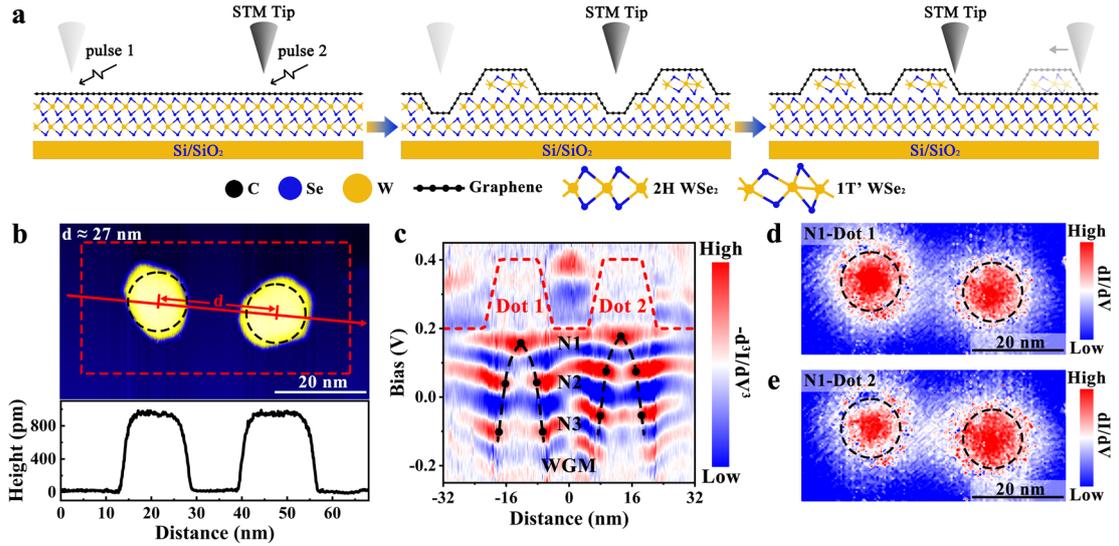

**Fig. 1 | Fabrication, manipulation and characterization of two GQDs. a,** Schematic of the strategy for fabricating and manipulating GQDs. **b,** Top: A representative STM image ($V_{bias} = 600$ mV, $I = 100$ pA) of the two GQDs with distance $d = 27$ nm. Bottom: a height profile across the centers of the two GQDs. **c,** The $-d^3I/dV^3$ STS map versus the spatial position along the red arrow in panel **b**. The red dashed line indicates Dirac point energies used in theory calculation. The black solid dots indicate the quasibound states via the WGM confinement. **d,e,** The energy-fixed $dI/dV$ maps of the red dotted box region in panel b at the lowest quasibound states (N1) of the dot 1 and dot 2.

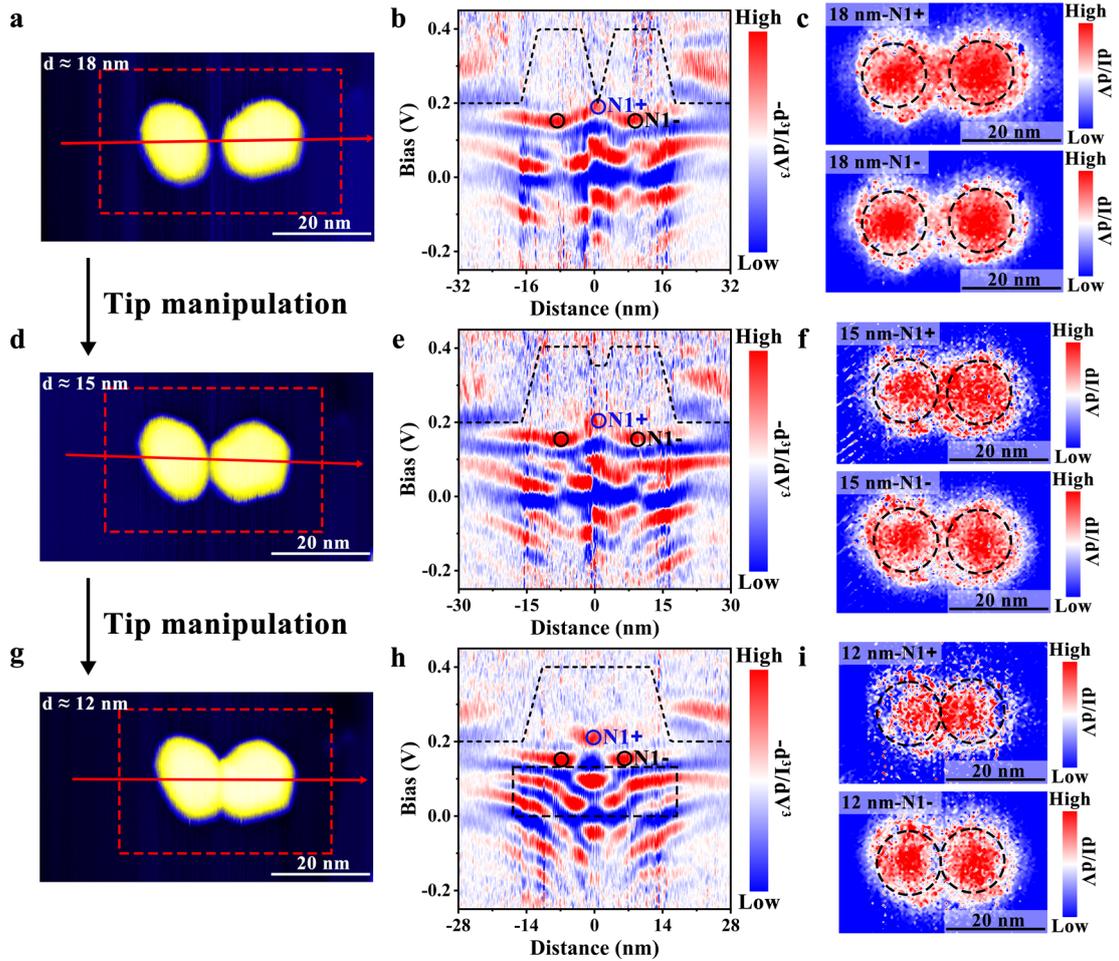

**Fig. 2 | Structures and electronic properties of two coupled GQDs with tunable distance. a,d,g,** From top panel to bottom panel: STM images of two coupled GQDs with decreasing distance. The distances are 18 nm (**a**), 15 nm (**d**) and 12 nm (**g**), respectively. The scanning parameters are all $V_{bias} = 600$ mV, $I = 100$ pA. **b,e,h,** The $-d^3I/dV^3$ STS maps versus the spatial position along the red arrow in panel **a**, **d** and **g**, respectively. The black dashed lines indicate Dirac point energies used in theory calculation. The blue and black circles indicate the bonding states (N1+) and antibonding states (N1-) of the two coupled GQDs. **c,f,i,** The energy-fixed $dI/dV$ mappings at the N1+ and N1- marked in panel **b**, **e** and **h**, respectively.

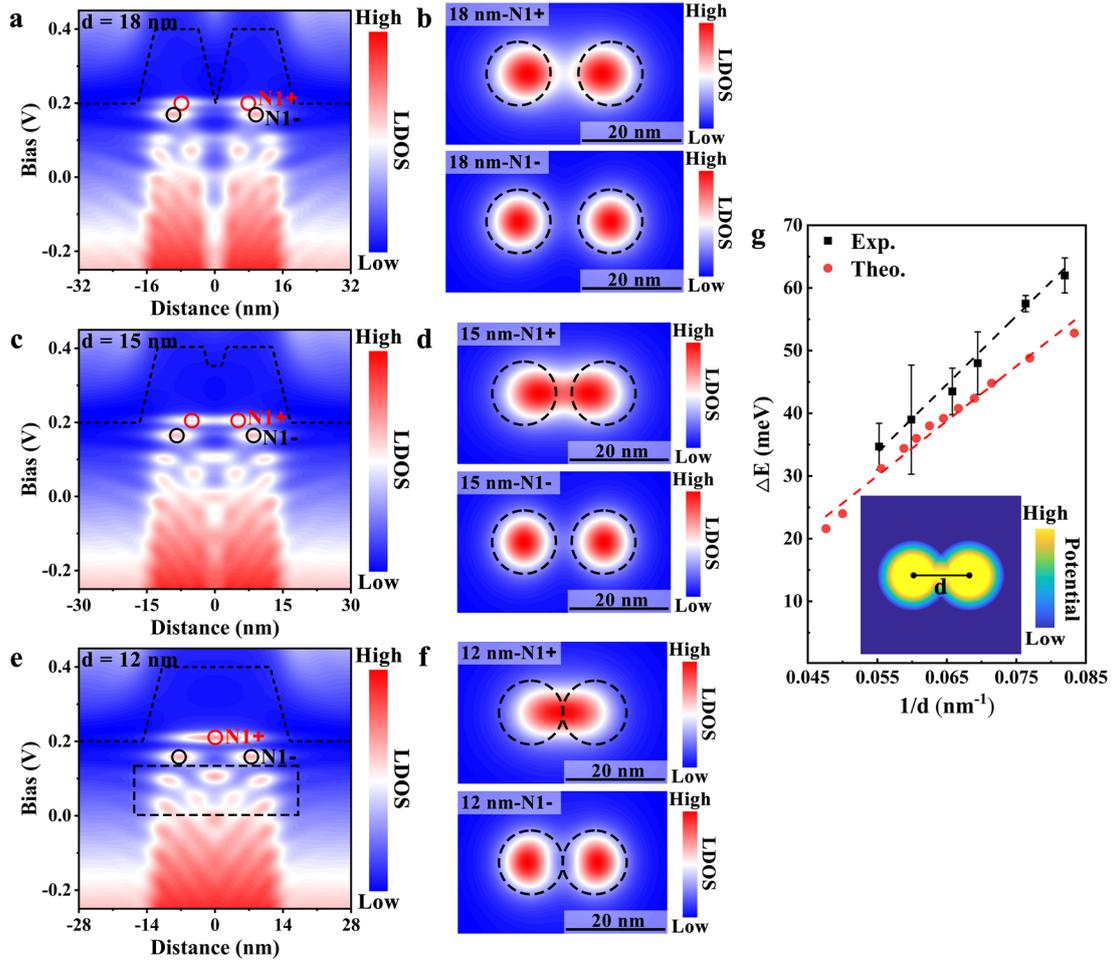

**Fig. 3 | The calculated electronic properties of two coupled GQDs with different distance. a,c,e,** From top panel to bottom panel, the calculated space-energy maps of the LDOS along the line crossing the centers of the two coupled GQDs in Fig. 2a, d and g, respectively. The black dashed lines indicate Dirac point energies used in theory calculation. The red and black circles indicate the bonding states (N1+) and antibonding states (N1-) of the two coupled GQDs. **b,d,f,** The calculated energy-fixed LDOS mappings at N1+ and N1- marked in panel **a, c** and **e**. **g,** The energy spacing $\Delta E$ of the N1+ and N1- as a function of $1/d$ between the two GQDs obtained in experiment and in theory. The inset shows the schematic of potential field distribution of two coupled GQDs.

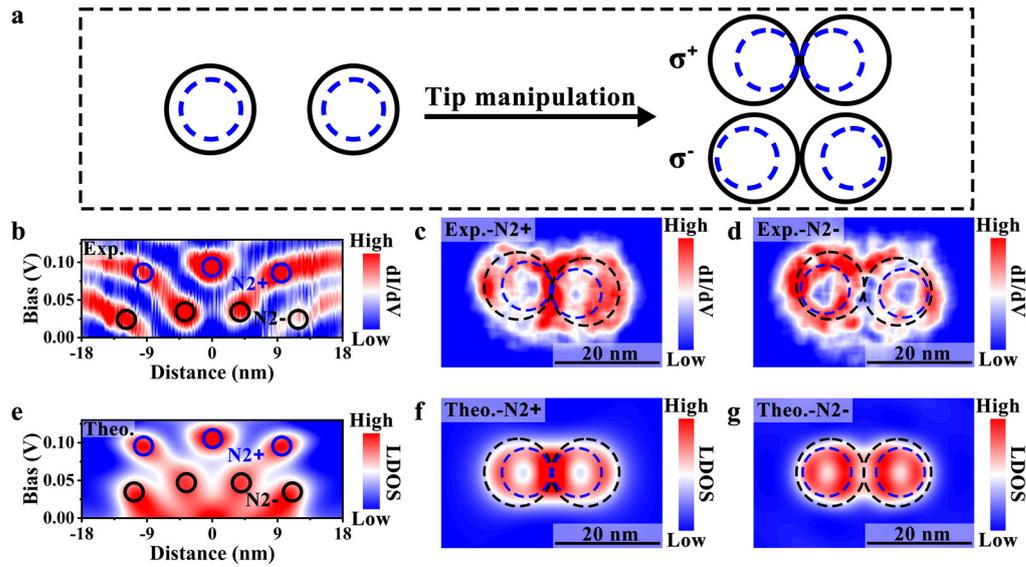

**Fig. 4 | Molecular states formed by the quasibound states with higher angular momenta. a,** Schematic of the behavior of the uncoupled (left) and coupled (right) quasibound states with higher angular momenta. The black solid circles represent the GQDs. The blue dashed circles represent one of the quasibound states with higher angular momenta. $\sigma^+$ and $\sigma^-$ represent the bonding and antibonding states of the quasibound states. **b,** Zoom-in $-d^3I/dV^3$ STS map of the black dashed rectangle in Fig. 2h. N2+ and N2- represent the bonding and antibonding states of the quasibound states with higher angular momenta, respectively. **c,d,** The energy-fixed $dI/dV$ maps at N2+ and N2- marked in panel **b**. **e,** Zoom-in calculated space-energy map of the LDOS of the black dashed rectangle in Fig. 3e. **f,g,** The calculated energy-fixed LDOS mappings at N2+ and N2- marked in panel **e**.

**Methods**

**Sample fabrication.** The graphene/WSe$_2$ heterostructure was fabricated by wet transfer technology of a graphene monolayer on mechanical-exfoliated thick 2H-phase WSe$_2$ sheets. Firstly, large area aligned graphene was grown on a 20 × 20 mm$^2$ polycrystalline copper (Cu) foil (Alfa Aesar, 99.8% purity, 25 μm thick) through chemical vapor deposition (CVD) method. The Cu foil was soaked in a solution of acetic acid and deionized water (volume ratio = 1:1) for 10 h. After that, the Cu foil was rinsed with alcohol and deionized water separately. The Cu foil was put in the CVD furnace annealing under an Ar/H2 flow (both at 50 sccm) at 1035 °C for 12 h at low pressure. Then the large-area aligned monolayer graphene was grown under a flow of Ar, H2 and CH4 (50 sccm, 50 sccm and 3 sccm separately) at 1035 °C for 30 minutes at low pressure. The system was cooled to room temperature slowly. Then, polymethyl methacrylate (PMMA) was uniformly coated on Cu foil with graphene monolayer. The Cu/graphene/PMMA film was put into ammonium persulfate solution until that the Cu foil was etched away. And the graphene/PMMA film was cleaned by deionized water for 6 times. The thick 2H-phase WSe$_2$ sheets were separated from the bulk crystal by traditional mechanical exfoliation technology and transfered to the Si/SiO2 substrate using polydimethylsiloxane (PDMS). Then, the Cu/graphene/PMMA film was put on the Si/SiO2 substrate with thick 2H-phase WSe$_2$ sheets. Lastly, after the film was dried, the PMMA was removed using acetone and the sample is cleaned by alcohol.

**STM/STS measurements.** The STM/STS measurements were performed in low-temperature (78 K) and ultrahigh-vacuum (~10$^{-10}$ Torr) scanning probe microscopes (USM-1500) from UNISOKU. The tips were obtained by electrochemical etching from a W (99.95%) alloy wire. The differential conductance (d$I$/d$V$) measurements were taken by a standard lock-in technique with an ac bias modulation of 5 mV and 793 Hz signal added to the tunneling bias.

**Theoretical model.** To numerically simulate the experiment results, we use the tight-binding model on the hexagonal lattice to describe the graphene, with the Hamiltonian being:

$$H = -\sum_{\langle ij \rangle} t_{ij}\, a_i^\dagger a_j + \sum_i V(\vec{r}_i)\, a_i^\dagger a_i \qquad (1)$$

where $t_{ij} = 3.2\,eV$ is the hopping energy between two nearest-neighbor sites i, j and is directly related to the Fermi velocity $v_F \approx 1.03 \times 10^6 m/s$ by $\hbar v_F = \frac{3}{2} t_{ij} a_{cc}$ ($a_{cc} = 0.142\,nm$ is the length of carbon-carbon bond). Here we only consider the nearest-neighbor bond $\langle i,j \rangle$. $a_i$ and $a_i^\dagger$ denote the annihilation and creation operators at site $i$, $\vec{r}_i = (x_i, y_i)$ is the position coordinate of the site $i$ relative to the origin, $V(\vec{r}_i)$ denotes the potential field on the graphene. Specially, for the system of the single GQD and coupled GQDs, we use $V_s(\vec{r})$ and $V_m(\vec{r})$ as shown in the main text, respectively. In the detail of simulations, a large hexagonal graphene flake is built with all the armchair edges (to avoid zigzag edge states at the low energy). The side length of the hexagon is 200 nm and the graphene flake system includes about 4 million carbon atoms, which is large enough to remove the finite size effect. We use the open source code package for numerical calculations: *Pybinding*[42]. Using the fast implementation of kernel polynomial method[43,44] in this package, we can quickly obtain the LDOS $\tilde{\rho}(\vec{r}_i)$ for each site $i$ with an appropriate energy broadening $\Gamma$. Furthermore, to better compared with the experiment results, we also introduce the space broadening $\lambda_s$ (simulating the broadening of the STM tip) to reformulate the LDOS at position $\vec{r}$: $\rho(\vec{r}) = \sum_i \tilde{\rho}(\vec{r}_i) e^{-\frac{|\vec{r}-\vec{r}_i|^2}{2\lambda_s^2}}$. In the calculations, we set $\lambda_s = 0.15 nm$ and $\Gamma = 0.01 eV$.